\documentclass[12pt,english]{article}
\usepackage[T1]{fontenc}
\usepackage[utf8x]{inputenc}
\usepackage[a4paper]{geometry}
\usepackage{babel}

\usepackage{array}
\usepackage{textcomp}
\usepackage{amsmath}
\usepackage{esint}
\usepackage[unicode=true, pdfusetitle,
 bookmarks=true,bookmarksnumbered=false,bookmarksopen=false,
 breaklinks=false,pdfborder={0 0 1},backref=false,colorlinks=false]
 {hyperref}

\makeatletter

\providecommand{\tabularnewline}{\\}

\newcommand{\lyxaddress}[1]{
\par {\raggedright #1
\vspace{1.4em}
\noindent\par}
}

\newcommand{\tref}[1]{\ref{#1}}

\makeatother

\begin{document}

\title{Electromagnetic source transformations and scalarization in stratified
gyrotropic media}

\author{P. De Visschere}

\maketitle

\lyxaddress{Ghent University, ELIS, Sint-Pietersnieuwstraat 41, B-9000 Gent\\
pdv@elis.ugent.be}
\begin{abstract}
It is known that with restrictions on the type of the constitutive
equations, Maxwell's equations in non-uniform media can sometimes
be reduced to two 2nd order differential equations for 2 scalar quantities
only. These results have previously been obtained in two quite different
ways. Either by a {}``scalarization of the sources'', where the
relevant scalar quantities are essentially vector potential components
and where the derivation was limited to isotropic media; or alternatively
by using the {}``scalar Hertz potentials'', and this method has
been applied to more general media. In this paper it is shown that
both methods are equivalent for gyrotropic media. We show that the
scalarization can be obtained by a combination of transformations
between electric and magnetic sources and gauge transformations. It
is shown that the method based on the vector potential, which previously
used a non-traditional definition of the vector potentials, can also
be obtained using the traditional definition provided a proper gauge
condition is applied and this method is then extended from isotropic
to gyrotropic media. It is shown that the 2 basic scalar Hertz potentials
occurring in the second method are invariant under the source scalarization
transformations of the first method and therefore are the natural
potentials for obtaining scalarization. Finally it is shown that both
methods are also equivalent with a much older third method based on
Hertz vectors.
\end{abstract}

\section{Introduction}

Usually problems in electromagnetics are reduced to solving a 2nd
order vectorial equation for either the electric field or for the
magnetic field \cite{John-D.-Joannopoulos:2008sf,Monk:2003rz}. However
for some problems, the radiation of a dipole in a stratified medium
being a good example, it is still useful to use a representation for
the fields in terms of auxiliary functions (vector and scalar potentials
or Hertz vector potentials) and then solving equations for these auxiliary
functions instead of for the fields directly. It is well-known that
for simple uniform electromagnetic media, Maxwell's equations can
in this way be reduced to 2 scalar uncoupled 2nd order differential
equations, corresponding with the TE/TM modes. For uniform media this
has been generalized to more complex (decomposable) media \cite{Olyslager:2001p1770}.
Also for non-uniform, usually stratified, media such a scalar decomposition
has been obtained.

One of the earliest systematic treatments of this {}``reduction problem''
was given by Nisbet \cite{Nisbet:1955p2231,Nisbet:1957p2232}. First
a general formulation in terms of 2 Hertz vector potentials was given
valid for non-uniform anisotropic media and then it was noticed that
considerable freedom exists for choosing these potentials since they
can be subjected to a gauge transformation. Using this freedom the
Hertz vectors could be reduced to single component vectors and conditions
were derived under which the resulting differential equations for
these components were of 2nd order, at least for isotropic media.
The same idea was later extended to anisotropic media \cite{MOHSEN:1973p2222,MOHSEN:1976p2225}.
The reduction to two scalar potentials was also extended to gyrotropic
media \cite{Przeziecki:1979p3377,WEIGLHOFER:1985p3541} and to even
more complicated media \cite{Weiglhofer:1999p4155}. Only rather recently
the case of a uniaxial medium was given explicitly \cite{Weiglhofer:2000p1762}.
Whereas in these earlier publications \cite{Nisbet:1957p2232,MOHSEN:1976p2225}
a general coordinate system was considered, the later extensions to
more complex media usually considered a cartesian coordinate system
only and the non-uniformity was limited to a stratification along
e.g. the symmetry axis of the uniaxial medium \cite{Weiglhofer:2000p1762}.
Although the 2 scalar potentials are referred to as {}``scalar Hertz
potentials'' the link with the Hertz vector potentials is not  obvious
anymore and for the stratified uniaxial medium the (initially 4) scalar
functions are instead defined by applying a Helmholtz-decomposition
to the electric and magnetic field components perpendicular to the
symmetry axis \cite{Weiglhofer:2000p1762}.

Subsequently Weiglhofer and Georgieva \cite{Georgieva:2002p1769,Weiglhofer:2003p1723}
arrived at the same scalar equations, at least for isotropic media,
following a completely different method, which rests mainly on the
so-called scalarization of sources. It was shown that arbitrary current
density distributions can be replaced by equivalent distributions
but oriented along a fixed direction. With a proper (unconventional)
choice of the vector potentials the latter could then also be scalarized
and the resulting equations are exactly the same as those found using
the scalar Hertz potentials. This remarkable correspondence was noticed
but no explanation was given \cite{Weiglhofer:2003p1723}.

The main purpose of this paper is to shed some light on this finding,
which can not be a coincidence. At the same time we will extend the
source scalarization method explained in \cite{Georgieva:2002p1769,Weiglhofer:2003p1723}
to more general (gyrotropic) anisotropic media. We believe that {}``source
scalarization'' can be understood best as an application of the well-known
equivalence between electric and magnetic charges and currents. This
is first presented in §~\ref{sec:Source-transformations}. In the
following sections the scalarization problem is solved using different
potentials. As in most of the referenced papers we consider only cartesian
coordinates. As medium a stratified initially uniaxial medium is considered
where the symmetry axis is perpendicular to the layers everywhere.
The latter condition is necessary to avoid  mixing between the longitudinal
and transverse field components by applying the constitutive equations.
In §~\ref{sec:Gyrotropic-media} the theory is extended to a stratified
gyrotropic medium. We will use $\overline{c}$ as a unit vector along
the symmetry axis and $c$ as the corresponding coordinate whereas
transversal vector components will be labeled by $\perp$, in particular
the transversal nabla operator will be written as $\nabla_{\perp}$.

\section{\label{sec:Source-transformations}Source transformations}

We use Maxwell's equations in the standard form including electric
and magnetic charge and current densities, where the latter are labeled
by a superscript star\begin{eqnarray}
\nabla\times\overline{E} & = & -\frac{\partial\overline{B}}{\partial t}-\overline{J}^{*}\label{eq:1}\\
\nabla\times\overline{H} & = & \frac{\partial\overline{D}}{\partial t}+\overline{J}\label{eq:2}\\
\nabla\cdot\overline{D} & = & \rho\label{eq:3}\\
\nabla\cdot\overline{B} & = & \rho^{*}\label{eq:4}\end{eqnarray}

The possibly position dependent constitutive properties of the medium
are given by $\overline{D}=\epsilon\cdot\overline{E}$ and $\overline{B}=\mu\cdot\overline{H}$.
We will write the electric charge and current densities in general
as\begin{eqnarray}
\overline{J} & = & \frac{\partial\overline{p}}{\partial t}+\nabla\times\overline{m}\\
\rho & = & -\nabla\cdot\overline{p}\end{eqnarray}

where $\overline{p}$ and $\overline{m}$ are either given polarization
and magnetization densities or must be considered as stream potentials
for given $\rho$ and $\overline{J}$ \cite{Nisbet:1957p2232}. In
either case $\overline{p}$ and $\overline{m}$ can be subjected to
a gauge transformation which leaves $\rho$ and $\overline{J}$ invariant
\cite{Nisbet:1957p2232}\begin{eqnarray}
\overline{p}' & = & \overline{p}+\nabla\times\overline{G}\label{eq:7}\\
\overline{m}' & = & \overline{m}-\frac{\partial\overline{G}}{\partial t}+\nabla g\label{eq:8}\end{eqnarray}

where $\overline{G},g$ are arbitrary functions. Since our aim is
to transform transversal sources into longitudinal ones, 2 possibilities
arise. With $\nabla_{\perp}g=-\overline{m}_{\perp}$ a transversal
magnetization is turned into a longitudinal one $\frac{\partial g}{\partial c}\overline{c}$.
And with a longitudinal $\overline{G}=G_{c}\overline{c}$ a transversal
polarization $\overline{p}_{\perp}=-\nabla_{\perp}G_{c}\times\overline{c}$
is turned into a longitudinal magnetization $-\frac{\partial G_{c}}{\partial t}\overline{c}$.

It is well-known that the polarization $\overline{p}$ and the magnetization
$\overline{m}$ can equally well be represented by magnetic charge
and current densities, which is most easily seen by rearranging the
terms in the Maxwell equations as follows\begin{eqnarray}
\nabla\times(\overline{E}+\epsilon^{-1}\cdot\overline{p}) & = & -\frac{\partial}{\partial t}(\overline{B}-\mu\cdot\overline{m})-\frac{\partial\mu\cdot\overline{m}}{\partial t}+\nabla\times\epsilon^{-1}\cdot\overline{p}\\
\nabla\times(\overline{H}-\overline{m}) & = & \frac{\partial}{\partial t}(\overline{D}+\overline{p})\\
\nabla\cdot(\overline{D}+\overline{p}) & = & 0\\
\nabla\cdot(\overline{B}-\mu\cdot\overline{m}) & = & -\nabla\cdot(\mu\cdot\overline{m})\end{eqnarray}

The equivalent magnetic sources are thus given by\begin{eqnarray}
\overline{J}^{*} & = & \frac{\partial\mu\cdot\overline{m}}{\partial t}-\nabla\times\epsilon^{-1}\cdot\overline{p}\\
\rho^{*} & = & -\nabla\cdot(\mu\cdot\overline{m})\end{eqnarray}

Contrary to the gauge transformation \eqref{eq:7}\eqref{eq:8}, in
this case the source transformation (from electrical charges to magnetic
charges) is accompanied by the following field transformations\begin{eqnarray}
\overline{E}' & = & \overline{E}+\epsilon^{-1}\cdot\overline{p}\label{eq:15}\\
\overline{B}' & = & \overline{B}-\mu\cdot\overline{m}\label{eq:16}\end{eqnarray}

In what follows we will label a polarization/magnetization density
which is represented by magnetic charges by a superscript star ($\overline{p}^{*},\overline{m}^{*}$).
We can thus freely exchange $\overline{p}$ (or $\overline{m}$) for
$\overline{p}^{*}$ (or $\overline{m}^{*}$) and vice versa as long
as $\overline{p}+\overline{p}^{*}$(or $\overline{m}+\overline{m}^{*}$)
remains invariant and as long as we take the field transformations
\eqref{eq:15}\eqref{eq:16} into account. These {}``magnetic''
stream potentials can also be subjected to a gauge transformation
\cite{Nisbet:1957p2232}\begin{eqnarray}
\epsilon^{-1}\cdot\overline{p}'^{*} & = & \epsilon^{-1}\cdot\overline{p}^{*}-\frac{\partial\overline{L}}{\partial t}+\nabla l\label{eq:17-1}\\
\mu\cdot\overline{m}'^{*} & = & \mu\cdot\overline{m}^{*}-\nabla\times\overline{L}\label{eq:18}\end{eqnarray}

for arbitrary $\overline{L},l$. With $l$ we can turn a transversal
polarization into a longitudinal one and with $\overline{L}=L_{c}\overline{c}$
we can turn a transversal magnetization into a longitudinal polarization.
For further reference the different representations are tabulated
in \tref{tab:sources}.

\begin{table}
\caption{\label{tab:sources}The equivalent source contributions due to (external)
polarization and magnetization. The starred quantities allow to make
a distinction between sources modeled with electric charges/currents
and those modeled with magnetic ones.}

\noindent \centering{}\begin{tabular}{|c||c|c|}
\hline 
 & polarization & magnetization\tabularnewline
\hline
\hline 
$\overline{J}$ & $\frac{\partial\overline{p}}{\partial t}$ & $\nabla\times\overline{m}$\tabularnewline
\hline 
$\rho$ & $-\nabla\cdot\overline{p}$ & \tabularnewline
\hline
\hline 
$\overline{J}^{*}$ & $-\nabla\times\epsilon^{-1}\cdot\overline{p}^{*}$ & $\frac{\partial\mu\cdot\overline{m}^{*}}{\partial t}$\tabularnewline
\hline 
$\rho^{*}$ &  & $-\nabla\cdot\mu\cdot\overline{m}^{*}$\tabularnewline
\hline
\end{tabular}
\end{table}

Using the electric/magnetic charge transformations and gauge transformations
if needed we can now scalarize an arbitrary current density%
\footnote{We will consider an electric current density, but the same method
can be applied to a magnetic current density.%
} along a fixed direction defined by the unit vector $\overline{c}$.
The goal of the scalarization process is to replace the current density
by equivalent electric and magnetic current densities parallel to
$\overline{c}$. For a stratified medium $\overline{c}$ is perpendicular
to the layers and as mentioned in the {}``Introduction'' this is
also the direction of the symmetry axis of the uniaxial medium. The
current density can always be written as\begin{equation}
\overline{J}=J_{c}\overline{c}+\nabla_{\perp}v\times\overline{c}+\nabla_{\perp}u\label{eq:17}\end{equation}

In the context of scalar Hertz potentials the functions $u,v$ are
known as auxiliary functions \cite{Weiglhofer:2000p1762} and they
can be found by solving the 2-dimensional potential problems\begin{eqnarray}
\nabla_{\perp}^{2}u & = & \nabla_{\perp}\cdot\overline{J}_{\perp}\label{eq:18-1}\\
\nabla_{\perp}^{2}v & = & -\overline{c}\cdot(\nabla_{\perp}\times\overline{J}_{\perp})\label{eq:19-1}\end{eqnarray}

Scalarization of the 2nd term in \eqref{eq:17} is straighforward,
since according to \tref{tab:sources} it can be attributed to a magnetization
$\overline{m}=v\overline{c}$ which can also be represented by a magnetic
current density along $\overline{c}$ \begin{equation}
\overline{J}^{*}=\frac{\partial\mu\cdot\overline{m}^{*}}{\partial t}=\mu_{//}\frac{\partial v}{\partial t}\overline{c}\end{equation}

where $\mu_{//}$ is the permeability along $\overline{c}$. This
transformation is accompanied by a field transformation according
to \eqref{eq:16}\begin{equation}
\overline{B}'=\overline{B}-\mu_{//}v\overline{c}\end{equation}

Scalarization of the last contribution in \eqref{eq:17} cannot be
obtained simply by transforming electric into magnetic sources. However
this transformation can always be combined with a gauge transformation.
The last contribution in \eqref{eq:17} can then be scalarized by
attributing the current density $\nabla_{\perp}u$ to a transversal
polarization density $\overline{p}=\nabla_{\perp}\int udt$ which,
as we have seen, can be turned into a longitudinal one in the {}``magnetic''
domain by a proper choice of $l$ in \eqref{eq:17-1} (with $\overline{L}=0$)
namely\begin{equation}
l=-\int\frac{u}{\epsilon_{\perp}}dt\label{eq:22}\end{equation}

where $\nabla_{\perp}\epsilon_{\perp}=0$ has been used. We then end
up with a scalarized current density\begin{equation}
J'_{c}=-\epsilon_{//}\frac{\partial}{\partial c}\left(\frac{u}{\epsilon_{\perp}}\right)\label{eq:23}\end{equation}

Due to the 2 electric/magnetic transformations preceding and following
the gauge transformation we must take into account a transformation
of the electric field according to \eqref{eq:15}\begin{equation}
\overline{E}'-\overline{E}=\nabla l=\nabla\left(\int\frac{u}{\epsilon_{\perp}}dt\right)\label{eq:24}\end{equation}

These results as well as similar ones for the magnetic current density
are tabulated in \tref{tab:transforms_e} and \tref{tab:transforms_m}.
To conclude this section we make 2 remarks
\begin{enumerate}
\item It can be proved that the scalarizations summarized in \tref{tab:transforms_e}
and \tref{tab:transforms_m} are unique;
\item Under these source transformations the total longitudinal current
densities ($J_{c}+\frac{\partial D_{c}}{\partial t}$ and $J_{c}^{*}+\frac{\partial B_{c}}{\partial t}$)
are invariant, since \begin{eqnarray}
J_{c}+\frac{\partial D_{c}}{\partial t} & = & \left(\nabla\times\overline{H}\right)\cdot\overline{c}\\
J_{c}^{*}+\frac{\partial B_{c}}{\partial t} & = & -\left(\nabla\times\overline{E}\right)\cdot\overline{c}\end{eqnarray}
and $\overline{H}'-\overline{H}$ and $\overline{E}'-\overline{E}$
are either parallel to $\overline{c}$ or equal to a gradient.
\end{enumerate}
\begin{table}
\caption{\label{tab:transforms_e}Using proper transformations between electric
and magnetic charge representations from \tref{tab:sources} and possibly
gauge transformations of the stream potentials, arbitrary transverse
electric current densities can be replaced by equivalent electric
or magnetic current densities along the symmetry direction $\overline{c}$.
Each transformation is also accompanied by a transformation of the
fields, shown in the last 2 rows.}

\noindent \centering{}\begin{tabular}{|>{\centering}p{2cm}||c|c|}
\hline 
initial current density & $\overline{J}=\nabla_{\perp}v\times\overline{c}$ & $\overline{J}=\nabla_{\perp}u$\tabularnewline
\hline
\hline 
initial stream potenial & $\overline{m}=v\overline{c}$ & $\overline{p}=\nabla_{\perp}\int udt$\tabularnewline
\hline 
final stream potential & $\overline{m}^{*}=v\overline{c}$ & $\overline{p}=-\epsilon_{//}\frac{\partial}{\partial c}\left(\int\frac{u}{\epsilon_{\perp}}dt\right)\overline{c}$\tabularnewline
\hline
\hline 
final current density & $\overline{J}^{*}=\mu_{//}\frac{\partial v}{\partial t}\overline{c}$ & $\overline{J}=-\epsilon_{//}\frac{\partial}{\partial c}\left(\frac{u}{\epsilon_{\perp}}\right)\overline{c}$\tabularnewline
\hline
\hline 
$\overline{E}'-\overline{E}$ &  & $\nabla\left(\int\frac{u}{\epsilon_{\perp}}dt\right)$\tabularnewline
\hline 
$\overline{B}'-\overline{B}$ & $-\mu_{//}v\overline{c}$ & \tabularnewline
\hline
\end{tabular}
\end{table}

\begin{table}
\caption{\label{tab:transforms_m}Using proper transformations between electric
and magnetic charge representations from \tref{tab:sources} and possibly
gauge transformations of the stream potentials, arbitrary transverse
magnetic current densities can be replaced by equivalent electric
or magnetic current densities along the symmetry direction $\overline{c}$.
Each transformation is also accompanied by a transformation of the
fields, shown in the last 2 rows.}

\noindent \centering{}\begin{tabular}{|>{\centering}p{2cm}||c|c|}
\hline 
initial current density & $\overline{J}^{*}=\nabla_{\perp}v^{*}\times\overline{c}$ & $\overline{J}^{*}=\nabla_{\perp}u^{*}$\tabularnewline
\hline
\hline 
initial stream potenial & $\overline{p}^{*}=-\epsilon_{//}v^{*}\overline{c}$ & $\overline{m}^{*}=\nabla_{\perp}\int\frac{u^{*}}{\mu_{\perp}}dt$,\tabularnewline
\hline 
final stream potential & $\overline{p}=-\epsilon_{//}v^{*}\overline{c}$ & $\overline{m}^{*}=-\frac{\partial}{\partial c}\left(\int\frac{u^{*}}{\mu_{\perp}}dt\right)\overline{c}$\tabularnewline
\hline
\hline 
final current density & $\overline{J}=-\epsilon_{//}\frac{\partial v^{*}}{\partial t}\overline{c}$ & $\overline{J}^{*}=-\mu_{//}\frac{\partial}{\partial c}\left(\frac{u^{*}}{\mu_{\perp}}\right)\overline{c}$\tabularnewline
\hline
\hline 
$\overline{E}'-\overline{E}$ & $v^{*}\overline{c}$ & \tabularnewline
\hline 
$\overline{B}'-\overline{B}$ &  & $\mu\cdot\nabla\left(\int\frac{u^{*}}{\mu_{\perp}}dt\right)$\tabularnewline
\hline
\end{tabular}
\end{table}

\section{\label{sec:Vector-potentials}Vector potentials}

We introduce conventional vector and scalar potentials $\overline{A},\phi$
and also comparable potentials $\overline{A}^{*},\phi^{*}$ for handling
the magnetic sources \begin{eqnarray}
\overline{B} & = & \nabla\times\overline{A}-\mu\cdot\frac{\partial\overline{A}^{*}}{\partial t}-\mu\cdot\nabla\phi^{*}\label{eq:27}\\
\overline{E} & = & -\frac{\partial\overline{A}}{\partial t}-\nabla\phi-\epsilon^{-1}\cdot\nabla\times\overline{A}^{*}\label{eq:28}\end{eqnarray}

Substitution into Maxwell's curl-equations gives initially\begin{eqnarray}
\mathcal{L}(\epsilon,\mu)\overline{A}+\epsilon\cdot\nabla\frac{\partial\phi}{\partial t} & = & \overline{J}\label{eq:29}\\
\mathcal{L}(\mu,\epsilon)\overline{A}^{*}+\mu\cdot\nabla\frac{\partial\phi^{*}}{\partial t} & = & \overline{J}^{*}\label{eq:30}\end{eqnarray}

where the operator $\mathcal{L}(\epsilon,\mu)$ is defined by\begin{equation}
\mathcal{L}(\epsilon,\mu)=\epsilon\cdot\frac{\partial^{2}}{\partial t^{2}}+\nabla\times\mu^{-1}\cdot\nabla\times\label{eq:47}\end{equation}

The scalar potentials are eliminated using gauge conditions. If these
gauge conditions do not mix-up the electric and magnetic quantities
then also the resulting equations will remain uncoupled. We first
mention the gauge conditions used by Nisbet \cite{Nisbet:1957p2232}\begin{eqnarray}
\nabla\cdot\epsilon\cdot\overline{A}+\alpha\frac{\partial\phi}{\partial t} & = & 0\label{eq:31}\\
\nabla\cdot\mu\cdot\overline{A}^{*}+\alpha^{*}\frac{\partial\phi^{*}}{\partial t} & = & 0\label{eq:32}\end{eqnarray}

where $\alpha,\alpha^{*}$ are scalars which can still be chosen.
As will become clear further on these gauge conditions allow scalarization
for a homogeneous medium only. This restriction is eliminated in \cite{Georgieva:2002p1769,Weiglhofer:2003p1723}
but only for an isotropic medium by using a different definition for
the potentials and using different gauge conditions. At first this
leads to equations for $\overline{A}$ and $\overline{A}^{*}$ which
are coupled but under the restrictions $\nabla_{\perp}\epsilon=0$
and $\nabla_{\perp}\mu=0$ these equations become uncoupled and scalarizable.
Although this method can be extended to uniaxial media we prefer to
stick with the conventional decompositions \eqref{eq:27}\eqref{eq:28}
and we will now derive gauge conditions which allow to scalarize \eqref{eq:29}\eqref{eq:30}
for a non-uniform uniaxial medium. 

We assume that the sources have already been scalarized so that in
\eqref{eq:29}\eqref{eq:30} only longitudinal current densities $J_{c},J_{c}^{*}$
occur. Splitting these equations into longitudinal and transversal
components the latter equations will only allow the null solution
for the transversal components of the vector potentials if their longitudinal
components do not occur in these transversal equations. These conditions
are easily found by assuming $\overline{A}=A_{c}\overline{c}$ and
equating the transversal component of the LHS of \eqref{eq:29} to
zero\begin{equation}
\nabla_{\perp}\left[\frac{\partial}{\partial c}\left(\frac{A_{c}}{\mu_{\perp}}\right)+\epsilon_{\perp}\frac{\partial\phi}{\partial t}\right]=0\label{eq:33-1}\end{equation}

and a similar {}``magnetic'' equation, which is obtained by replacing
unstarred quantities by starred ones and by switching the roles of
$\epsilon$ and $\mu$. Obviously using the gauge condition \eqref{eq:31},
this condition can only be met if $\epsilon_{\perp},\mu_{\perp}$
do not depend on the longitudinal coordinate \emph{c} and then only
by choosing $\alpha=\epsilon_{//}\epsilon_{\perp}\mu_{\perp}$. For
a more general result we must instead choose the following gauge conditions\begin{eqnarray}
\nabla\cdot\mu_{\perp}^{-1}\overline{A}+\epsilon_{\perp}\frac{\partial\phi}{\partial t} & = & 0\\
\nabla\cdot\epsilon_{\perp}^{-1}\overline{A}^{*}+\mu_{\perp}\frac{\partial\phi^{*}}{\partial t} & = & 0\end{eqnarray}

From the longitudinal components of \eqref{eq:29}\eqref{eq:30} and
these gauge conditions we then find the final scalarized equations

\begin{eqnarray}
\mathcal{L}_{s}(\epsilon,\mu)\left(\frac{{A'}_{c}}{\mu_{\perp}}\right) & = & -{J'}_{c}\label{eq:39}\\
\mathcal{L}_{s}(\mu,\epsilon)\left(\frac{{A'}_{c}^{*}}{\epsilon_{\perp}}\right) & = & -{J'}_{c}^{*}\label{eq:40}\end{eqnarray}
where scalarized quantities are now explicitly marked by an accent
and with the (scalar) wave operator defined by \begin{equation}
\mathcal{L}_{s}(\epsilon,\mu)=\nabla_{\perp}^{2}+\epsilon_{//}\frac{\partial}{\partial c}\epsilon_{\perp}^{-1}\frac{\partial}{\partial c}-\epsilon_{//}\mu_{\perp}\frac{\partial^{2}}{\partial t^{2}}\label{eq:scalar_operator}\end{equation}

Taking into account \eqref{eq:27}\eqref{eq:28} and the field transformations
due to the prior source scalarization, the fields are given by\begin{eqnarray}
\overline{H} & = & \left(v-\frac{\partial{A'}_{c}^{*}}{\partial t}\right)\overline{c}+\mu_{\perp}^{-1}\nabla_{\perp}{A'}_{c}\times\overline{c}-\nabla\left(\mu_{\perp}^{-1}\int\left[u^{*}-\frac{\partial}{\partial c}\left(\frac{{A'}_{c}^{*}}{\epsilon_{\perp}}\right)\right]dt\right)\label{eq:42-1}\\
\overline{E} & = & -\left(v^{*}+\frac{\partial{A'}_{c}}{\partial t}\right)\overline{c}-\epsilon_{\perp}^{-1}\nabla_{\perp}{A'}_{c}^{*}\times\overline{c}-\nabla\left(\epsilon_{\perp}^{-1}\int\left[u-\frac{\partial}{\partial c}\left(\frac{{A'}_{c}}{\mu_{\perp}}\right)\right]dt\right)\label{eq:43-1}\end{eqnarray}

\section{\label{sec:Hertz-vectors}Hertz vectors}

The main idea behind the use of Hertz vectors is to introduce an additional
differentiation in such a way that the new potentials (=~Hertz vectors)
are governed by equations with the polarization/magnetization (or
the stream potentials) as sources instead of the current densities.
Since for such Hertz vectors the starred/unstarred stream potentials
are equivalent we can eliminate e.g. the {}``magnetic charge'' sources
from the start and we can use instead of \eqref{eq:27}\eqref{eq:28}
the simpler decompositions\begin{eqnarray}
\overline{B}+\mu\cdot\overline{m}^{*} & = & \nabla\times\overline{A}\label{eq:42}\\
\overline{E}-\epsilon^{-1}\cdot\overline{p}^{*} & = & -\frac{\partial\overline{A}}{\partial t}-\nabla\phi\label{eq:43}\end{eqnarray}

still complying with \eqref{eq:1} and \eqref{eq:4}. The Hertz vector
equations are obtained most easily in the \emph{temporal gauge} \cite{Jackson:2002p3555}
($\phi=0$) and we then define the Hertz vectors $\overline{\Pi}_{e},\overline{\Pi}_{m}$
following \cite{Nisbet:1957p2232}\begin{equation}
\overline{A}=\frac{\partial\overline{\Pi}_{e}}{\partial t}+\epsilon^{-1}\cdot\nabla\times\overline{\Pi}_{m}\label{eq:44}\end{equation}

After substituting these equations in \eqref{eq:2}\eqref{eq:3} and
following a standard procedure \cite{Jackson:1999rz} we find the
equations\begin{eqnarray}
\mathcal{L}(\epsilon,\mu)\overline{\Pi}_{e} & = & \overline{p}+\overline{p}^{*}+\epsilon\cdot\nabla\psi_{e}\label{eq:45}\\
\mathcal{L}(\mu,\epsilon)\overline{\Pi}_{m} & = & \mu\cdot\left(\overline{m}+\overline{m}^{*}\right)+\mu\cdot\nabla\psi_{m}\label{eq:46}\end{eqnarray}

where the operator $\mathcal{L}(\epsilon,\mu)$ has been defined in
\eqref{eq:47}. The functions $\psi_{e},\psi_{m}$ can be chosen arbitrarely
and are in fact redundant due to the gauge transformations \eqref{eq:7}\eqref{eq:8}\eqref{eq:17-1}\eqref{eq:18}.
Substituting \eqref{eq:44} in \eqref{eq:42} and \eqref{eq:43} and
replacing the 2nd order time derivative using \eqref{eq:45} and \eqref{eq:47}
we find the following symmetric expressions for the fields \begin{eqnarray}
\overline{B}+\mu\cdot\overline{m}^{*} & = & \nabla\times\frac{\partial\overline{\Pi}_{e}}{\partial t}+\nabla\times\epsilon^{-1}\cdot\nabla\times\overline{\Pi}_{m}\label{eq:48}\\
\epsilon\cdot\overline{E}+\overline{p} & = & \nabla\times\mu^{-1}\cdot\nabla\times\overline{\Pi}_{e}-\nabla\times\frac{\partial\overline{\Pi}_{m}}{\partial t}\label{eq:49}\end{eqnarray}

Using the gauge transformations \eqref{eq:7}\eqref{eq:8}\eqref{eq:17-1}\eqref{eq:18}
one can try to simplify the equations \eqref{eq:45}\eqref{eq:46}.
One possibility is to eliminate the magnetizations ($\overline{m}+\overline{m}^{*}\rightarrow0$)
so that also $\overline{\Pi}_{m}\rightarrow0$. This is the essence
of the 1-Hertz-vector method followed by Sein \cite{SEIN:1989p2236},
in this case limited to a uniform isotropic medium and (therefore)
using the conventional Lorentz gauge. Another possibility is to scalarize
the stream functions so that $\overline{p}+\overline{p}^{*}\rightarrow{p'}_{c}\overline{c}$
and $\overline{m}+\overline{m}^{*}\rightarrow{m'}_{c}^{*}\overline{c}$.
For these scalarized sources it is now possible to choose the (gauge)
functions $\psi_{e},\psi_{m}$ in \eqref{eq:45}\eqref{eq:46} in
such a way that $\overline{\Pi}_{e},\overline{\Pi}_{m}$ are also
scalarized and thus have only components along $\overline{c}$. To
that end we assume $\overline{\Pi}_{e/m}=\Pi_{(e/m)c}\overline{c}$
and collect the transversal components of e.g. \eqref{eq:45}

\begin{eqnarray}
\frac{\partial}{\partial c}\left[\mu_{\perp}^{-1}\left(\nabla_{\perp}\Pi_{ec}\right)\right] & = & \epsilon_{\perp}\nabla_{\perp}\psi_{e}\label{eq:50}\end{eqnarray}

The transversal components of the vector potentials will then vanish
if we choose\begin{equation}
\psi_{e}=\frac{1}{\epsilon_{\perp}}\frac{\partial}{\partial c}\frac{\Pi_{ec}}{\mu_{\perp}}\label{eq:51}\end{equation}

The longitudinal part of the same equation is given by\begin{equation}
\epsilon_{//}\frac{\partial^{2}\Pi_{ec}}{\partial t^{2}}-\nabla_{\perp}^{2}\frac{\Pi_{ec}}{\mu_{\perp}}={p'}_{c}+\epsilon_{//}\frac{\partial\psi_{e}}{\partial c}\label{eq:52}\end{equation}

and using \eqref{eq:51} we obtain\begin{equation}
\mathcal{L}_{s}(\epsilon,\mu)\left(\frac{{\Pi'}_{ec}}{\mu_{\perp}}\right)=-{p'}_{c}\label{eq:53}\end{equation}

where the scalar operator $\mathcal{L}_{s}$ has been defined in \eqref{eq:scalar_operator}.
A similar equation can be found starting from \eqref{eq:46}\begin{equation}
\mathcal{L}_{s}(\mu,\epsilon)\left(\frac{{\Pi'}_{mc}}{\epsilon_{\perp}}\right)=-\mu_{//}{m'}_{c}^{*}\label{eq:54}\end{equation}

Unlike the scalarized current densities, which are unique, the scalarized
stream potentials and the corresponding Hertz vector components are
not unique, since they can always be subjected to a gauge transformation.
However we can in particular choose the scalarized stream potentials
as follows\begin{equation}
{J'}_{c}=\frac{\partial{p'}_{c}}{\partial t}\qquad{J'}_{c}^{*}=\mu_{//}\frac{\partial{m'}_{c}^{*}}{\partial t}\label{eq:55}\end{equation}

where ${J'}_{c}$ and ${J'}_{c}^{*}$ are the (unique) scalarized
current densities. Comparing \eqref{eq:53}\eqref{eq:54} with \eqref{eq:39}\eqref{eq:40}
we conclude that in that case \begin{equation}
{A'}_{c}=\frac{\partial{\Pi'}_{ec}}{\partial t}\qquad{A'}_{c}^{*}=\frac{\partial{\Pi'}_{mc}}{\partial t}\end{equation}

This correspondence can also be  checked by comparing the field expressions
\eqref{eq:48}\eqref{eq:49} with \eqref{eq:42-1}\eqref{eq:43-1}
where in \eqref{eq:48}\eqref{eq:49} one should also take into account
the field transformations due to the scalarization of the current
densities (see \tref{tab:transforms_e} and \tref{tab:transforms_m}).

\section{\label{sec:Scalar-Hertz-potentials}Scalar Hertz potentials}

The {}``scalar Hertz potential'' formulation introduced by Weiglhofer
\cite{Weiglhofer:2000p1762} starts by decomposing the fields and
the equations into transversal and longitudinal components and parts.
From these equations the longitudinal components $E_{c},H_{c}$ can
be eliminated leaving 4 equations for the 4 unknown transversal components.
Up to this point the method is identical to the 4$\times$4 matrix
method used for solving Maxwell's equations in stratified media \cite{BERREMAN:1972p971}.
However for dealing with the source terms and unlike the 4$\times$4
matrix method, the transversal field components are then expressed
using scalar (Hertz) potential functions\begin{eqnarray}
\overline{E}_{\perp} & = & \nabla_{\perp}\Phi+\nabla_{\perp}\times\Theta\overline{c}\label{eq:57}\\
\overline{H}_{\perp} & = & \nabla_{\perp}\Pi+\nabla_{\perp}\times\Psi\overline{c}\label{eq:58}\end{eqnarray}

In what follows we will give a compact derivation of the scalar Hertz
potential equations, following \cite{Weiglhofer:2000p1762}. Splitting
Maxwell's curl-equations parallel and perpendicular to $\overline{c}$
we obtain\begin{eqnarray}
\overline{c}\cdot(\nabla_{\perp}\times\overline{H}_{\perp}) & = & J_{c}+\frac{\partial D_{c}}{\partial t}\label{eq:59}\\
\overline{c}\cdot(\nabla_{\perp}\times\overline{E}_{\perp}) & = & -(J_{c}^{*}+\frac{\partial B_{c}}{\partial t})\label{eq:60}\end{eqnarray}

and\begin{eqnarray}
\nabla_{\perp}H_{c}-\frac{\partial\overline{H}_{\perp}}{\partial c} & = & \overline{c}\times(\overline{J}_{\perp}+\frac{\partial\overline{D}_{\perp}}{\partial t})\label{eq:61}\\
\nabla_{\perp}E_{c}-\frac{\partial\overline{E}_{\perp}}{\partial c} & = & -\overline{c}\times(\overline{J}_{\perp}^{*}+\frac{\partial\overline{B}_{\perp}}{\partial t})\label{eq:62}\end{eqnarray}

Taking the cross-product with $\overline{c}$, the latter 2 equations
become\begin{eqnarray}
\overline{c}\times\nabla_{\perp}H_{c}-\frac{\partial\left(\overline{c}\times\overline{H}_{\perp}\right)}{\partial c} & = & -\overline{J}_{\perp}-\frac{\partial\overline{D}_{\perp}}{\partial t}\label{eq:63}\\
\overline{c}\times\nabla_{\perp}E_{c}-\frac{\partial\left(\overline{c}\times\overline{E}_{\perp}\right)}{\partial c} & = & \overline{J}_{\perp}^{*}+\frac{\partial\overline{B}_{\perp}}{\partial t}\label{eq:64}\end{eqnarray}

The 4$\times$4 matrix method is based on \eqref{eq:62} and \eqref{eq:63},
without the source terms, and where \eqref{eq:59} and \eqref{eq:60}
are used for eliminating $E_{c}$ and $H_{c}$, after inserting the
constitutive equations\begin{eqnarray}
D_{c} & = & \epsilon_{//}E_{c}\qquad\overline{D}_{\perp}=\epsilon_{\bot}\overline{E}_{\perp}\label{eq:65}\\
B_{c} & = & \mu_{//}H_{c}\qquad\overline{B}_{\perp}=\mu_{\bot}\overline{H}_{\perp}\label{eq:66}\end{eqnarray}

Weiglhofer \cite{Weiglhofer:2000p1762} deals with the source terms
by operating with $\nabla_{\perp}\cdot$ on the 4 equations \eqref{eq:61}-\eqref{eq:64}
and by introducing the auxiliary functions already defined in \eqref{eq:18-1}
and \eqref{eq:19-1}. Using the constitutive equations \eqref{eq:65}\eqref{eq:66}
and the decompositions \eqref{eq:57}\eqref{eq:58} all terms then
contain the laplacian $\nabla_{\perp}^{2}$ which can be dropped yielding\begin{eqnarray}
H_{c}-\frac{\partial\Pi}{\partial c}-\epsilon_{\bot}\frac{\partial\Theta}{\partial t} & = & v\label{eq:67}\\
E_{c}-\frac{\partial\Phi}{\partial c}+\mu_{\bot}\frac{\partial\Psi}{\partial t} & = & -v^{*}\label{eq:68}\\
\epsilon_{\bot}\frac{\partial\Phi}{\partial t}-\frac{\partial\Psi}{\partial c} & = & -u\label{eq:69}\\
\mu_{\bot}\frac{\partial\Pi}{\partial t}+\frac{\partial\Theta}{\partial c} & = & -u^{*}\label{eq:70}\end{eqnarray}

On the other hand the longitudinal equations \eqref{eq:59}\eqref{eq:60}
become\begin{eqnarray}
-\nabla_{\perp}^{2}\Psi & = & J_{c}+\epsilon_{//}\frac{\partial E_{c}}{\partial t}\label{eq:71}\\
\nabla_{\perp}^{2}\Theta & = & J_{c}^{*}+\mu_{//}\frac{\partial H_{c}}{\partial t}\label{eq:72}\end{eqnarray}

Finally $\partial E_{c}/\partial t$ and $\partial H_{c}/\partial t$
can be calculated from \eqref{eq:67}-\eqref{eq:70} as a function
of $\Psi,\Theta$ only and when substituted in \eqref{eq:71}\eqref{eq:72}
one obtains two uncoupled 2nd order equations in $\Psi,\Theta$. We
remind the reader that the RHSs of \eqref{eq:59}\eqref{eq:60} and
thus also of \eqref{eq:71}\eqref{eq:72} are invariant under the
source scalarization transformations shown in \tref{tab:transforms_e}
and \tref{tab:transforms_m}, meaning that $\nabla_{\perp}\times\overline{E}_{\perp}$
and $\nabla_{\perp}\times\overline{H}_{\perp}$and also $\Psi,\Theta$
are invariant under these transformations. Therefore the scalar potentials
$\Psi,\Theta$ can only depend on the unique scalarized current densities
${J'}_{c}$ and ${J'}_{c}^{*}$. Whereas using the vector potential
method in §~\ref{sec:Vector-potentials} or the Hertz vector method
in §~\ref{sec:Hertz-vectors} scalarization could only be obtained
with some effort by applying appropriate source and gauge transformations,
the scalar Hertz potentials $\Psi$ and $\Theta$ are the natural
potentials for obtaining scalarization since they are invariant under
the required transformations. The final equations can then also be
obtained immediately by making the RHSs of \eqref{eq:67}-\eqref{eq:70}
zero and replacing $J_{c},J_{c}^{*}$ in \eqref{eq:71}\eqref{eq:72}
by the scalarized versions ${J'}_{c},{J'}_{c}^{*}$ leading to\begin{eqnarray}
\mathcal{L}_{s}(\epsilon,\mu)\left(\Psi\right) & = & -{J'}_{c}\label{eq:73}\\
\mathcal{L}_{s}(\mu,\epsilon)\left(-\Theta\right) & = & -{J'}_{c}^{*}\label{eq:74}\end{eqnarray}

However note that $E_{c},H_{c}$ and $\Phi,\Pi$ are not invariant.
From \eqref{eq:69}\eqref{eq:70} we find\begin{eqnarray}
\Phi & = & \epsilon_{\perp}^{-1}\int\left(-u+\frac{\partial\Psi}{\partial c}\right)dt\\
\Pi & = & -\mu_{\perp}^{-1}\int\left(u^{*}+\frac{\partial\Theta}{\partial c}\right)dt\end{eqnarray}

and subsequently from \eqref{eq:67}\eqref{eq:68}\begin{eqnarray}
E_{c} & = & -v^{*}+\frac{\partial\Phi}{\partial c}-\mu_{\perp}\frac{\partial\Psi}{\partial t}\\
H_{c} & = & v+\frac{\partial\Pi}{\partial c}+\epsilon_{\perp}\frac{\partial\Theta}{\partial t}\end{eqnarray}

The total fields can then be written as\begin{eqnarray}
\overline{E} & = & -(v^{*}+\mu_{\perp}\frac{\partial\Psi}{\partial t})\overline{c}+\nabla_{\perp}\Theta\times\overline{c}+\nabla\Phi\label{eq:79}\\
\overline{H} & = & (v+\epsilon_{\perp}\frac{\partial\Theta}{\partial t})\overline{c}+\nabla_{\perp}\Psi\times\overline{c}+\nabla\Pi\label{eq:80}\end{eqnarray}

These expressions confirm the field transformations in \tref{tab:transforms_e}
and \tref{tab:transforms_m} and they correspond term for term with
the expressions in \eqref{eq:42-1}\eqref{eq:43-1} with the correspondence\begin{eqnarray}
\Psi=\frac{{A'}_{c}}{\mu_{\perp}}=\frac{\partial}{\partial t}\frac{{\Pi'}_{ec}}{\mu_{\perp}} & \qquad & -\Theta=\frac{{A'}_{c}^{*}}{\epsilon_{\perp}}=\frac{\partial}{\partial t}\frac{{\Pi'}_{mc}}{\epsilon_{\perp}}\label{eq:81}\\
\Phi=-\phi-\int\frac{u}{\epsilon_{\perp}}dt & \qquad & \Pi=-\phi^{*}-\int\frac{u^{*}}{\mu_{\perp}}dt\label{eq:82}\end{eqnarray}

which was already (partially) apparent from \eqref{eq:73}\eqref{eq:74}
and \eqref{eq:39}\eqref{eq:40}.

\section{\label{sec:Gyrotropic-media}Gyrotropic media}

For a gyrotropic medium the transversal dielectric tensor is given
by \cite{Przeziecki:1979p3377}\begin{eqnarray}
\epsilon_{\perp} & = & \left[\begin{array}{cc}
\epsilon_{\perp} & j\epsilon'_{\perp}\\
-j\epsilon'_{\perp} & \epsilon_{\perp}\end{array}\right]=\epsilon_{\perp}I+j\epsilon'_{\perp}\overline{c}\times I\end{eqnarray}

with a similar expression for the permittivity tensor.

Since the scalarization of the solenoidal parts of $\overline{J}_{\perp}$
and $\overline{J}'_{\perp}$ does not involve these transversal constitutive
tensors no changes are needed here. However for the irrotational parts
the reasoning leading to \eqref{eq:22}\eqref{eq:23} and \eqref{eq:24}
must be extended with additional terms. Introducing the gauge function
$G_{c}$ in the {}``electric'' domain \eqref{eq:7}\eqref{eq:8}
and as before $l$ in the {}``magnetic'' domain \eqref{eq:17-1}
we obtain the transformed stream potentials\begin{eqnarray}
\overline{p} & = & \nabla_{\perp}\int udt+\epsilon\cdot\nabla l+\nabla_{\perp}G_{c}\times\overline{c}\label{eq:89}\\
\overline{m}^{*} & = & -\frac{\partial G_{c}}{\partial t}\overline{c}\label{eq:90}\end{eqnarray}

If $\epsilon'_{\perp}\neq0$ then the 2nd term on the RHS of \eqref{eq:89}
contains an extra term which can be compensated by the last term if\begin{equation}
G_{c}=j\epsilon'_{\perp}l\end{equation}

In this way the extra transversal term in the polarization is transformed
into a longitudinal magnetization, which can be represented by a longitudinal
magnetic current density\begin{equation}
{J'}_{c}^{*}=-\mu_{//}\frac{\partial^{2}G_{c}}{\partial t^{2}}\end{equation}

Since \eqref{eq:22}\eqref{eq:23} and \eqref{eq:24} remain valid
it suffices thus to add a magnetic current density\begin{equation}
{J'}_{c}^{*}=j\frac{\epsilon'_{\perp}}{\epsilon_{\perp}}\mu_{//}\frac{\partial u}{\partial t}\end{equation}

and due to the electric/magnetic switch ($\overline{m}\rightarrow\overline{m}^{*})$
in \eqref{eq:90} we must also add a matching field transformation\begin{equation}
\overline{B}'-\overline{B}=\mu_{//}\frac{\partial G_{c}}{\partial t}\overline{c}=-j\frac{\epsilon'_{\perp}}{\epsilon_{\perp}}\mu_{//}u\overline{c}\end{equation}

Again $J{}_{c}+\frac{\partial B_{c}}{\partial t}$ remains invariant
for this additional transformation and therefore the formulation using
the \emph{scalar Hertz potentials} should still automatically lead
to scalarized equations as has been shown in \cite{WEIGLHOFER:1985p3541}.

Regarding the formulation using \emph{Hertz vectors} extra (transversal)
terms in $j\epsilon'_{\perp}$ will also occur in \eqref{eq:50}.
However the functions $\psi_{e/m}$ in \eqref{eq:45}\eqref{eq:46}
are only  part of the gauge transformation and more in general we
can also add terms in $\overline{G}$ respectively in $\overline{L}$
as in \eqref{eq:7}\eqref{eq:8} and \eqref{eq:17-1}\eqref{eq:18}
to the RHS of \eqref{eq:45}\eqref{eq:46}. With in particular $\overline{G}=G_{c}\overline{c}$
(and $\overline{L}=L_{c}\overline{c}$) we then obtain instead of
\eqref{eq:50} condition\begin{eqnarray}
\frac{\partial}{\partial c}\left(\frac{\mu_{\perp}}{\left|\mu_{\perp}\right|}\nabla_{\perp}\Pi_{ec}\right)-\overline{c}\times\frac{\partial}{\partial c}\left(\frac{j\mu'_{\perp}}{\left|\mu_{\perp}\right|}\nabla_{\perp}\Pi_{ec}\right) & = & \epsilon_{\perp}\nabla_{\perp}\psi_{e}\nonumber \\
 &  & +j\epsilon'_{\perp}\overline{c}\times\nabla_{\perp}\psi_{e}+\nabla G_{c}\times\overline{c}\label{eq:95}\end{eqnarray}

where $\left|\mu_{\perp}\right|=\det\,\mu_{\perp}=\mu_{\perp}^{2}-{\mu'}_{\perp}^{2}$.
This condition is fulfilled by choosing\begin{eqnarray}
\psi_{e} & = & \epsilon_{\perp}^{-1}\frac{\partial}{\partial c}\left(\frac{\mu_{\perp}}{\left|\mu_{\perp}\right|}\Pi_{ec}\right)\\
G_{c} & = & j\left[\frac{\partial}{\partial c}\frac{{\mu'}_{\perp}}{\mu_{\perp}}+\frac{{\epsilon'}_{\perp}}{\epsilon_{\perp}}\frac{\partial}{\partial c}\right]\frac{\mu_{\perp}}{\left|\mu_{\perp}\right|}\Pi_{ec}\label{eq:97}\end{eqnarray}

The longitudinal equation \eqref{eq:52} is replaced by\begin{equation}
\epsilon_{//}\frac{\partial^{2}\Pi_{ec}}{\partial t^{2}}-\nabla_{\perp}^{2}\left(\frac{\mu_{\perp}}{\left|\mu_{\perp}\right|}\Pi_{ec}\right)={p'}_{c}+\epsilon_{//}\left(\frac{\partial\psi_{e}}{\partial c}-\frac{\partial L_{c}}{\partial t}\right)\end{equation}

where $-L_{c}$ is given by a similar expression as in \eqref{eq:97}.
We notice that except for the replacement of $\mu_{\perp}^{-1}$ by
$\mu_{\perp}/\left|\mu_{\perp}\right|$ the main change is the occurence
of a cross-coupling term due to the additional gauge functions $G_{c},L_{c}$
which indeed mix between electric and magnetic stream functions. If
we replace \eqref{eq:scalar_operator} by the more general expression\begin{equation}
\mathcal{L}_{s}(\epsilon,\mu)=\nabla_{\perp}^{2}+\epsilon_{//}\frac{\partial}{\partial c}\epsilon_{\perp}^{-1}\frac{\partial}{\partial c}-\epsilon_{//}\frac{\left|\mu_{\perp}\right|}{\mu_{\perp}}\frac{\partial^{2}}{\partial t^{2}}\end{equation}

then the scalarized equations for a gyrotropic medium (omitting the
accents) are given by\begin{eqnarray}
\mathcal{L}_{s}(\epsilon,\mu)\Sigma_{e}+j\epsilon_{//}\left[\frac{\partial}{\partial c}\frac{{\epsilon'}_{\perp}}{\epsilon_{\perp}}+\frac{{\mu'}_{\perp}}{\mu_{\perp}}\frac{\partial}{\partial c}\right]\frac{\partial\Sigma_{m}}{\partial t}= & - & p_{c}\label{eq:100}\\
\mathcal{L}_{s}(\mu,\epsilon)\Sigma_{m}-j\mu_{//}\left[\frac{\partial}{\partial c}\frac{{\mu'}_{\perp}}{\mu_{\perp}}+\frac{{\epsilon'}_{\perp}}{\epsilon_{\perp}}\frac{\partial}{\partial c}\right]\frac{\partial\Sigma_{e}}{\partial t}= & - & \mu_{//}m_{c}^{*}\label{eq:101}\end{eqnarray}

where $\Sigma_{e}=\frac{\mu_{\perp}}{\left|\mu_{\perp}\right|}\Pi_{ec}$
and $\Sigma_{m}=\frac{\epsilon_{\perp}}{\left|\epsilon_{\perp}\right|}\Pi_{mc}$
and as for the uniaxial case these Hertz vector components are equivalent
with the scalar Hertz potentials with $\Psi=\partial\Sigma_{e}/\partial t$
and $-\Theta=\partial\Sigma_{m}/\partial t$.

Finally we consider the formulation using ordinary but standard \emph{vector
potentials}. Comparing \eqref{eq:29}\eqref{eq:30} with \eqref{eq:45}\eqref{eq:46}
the close resemblance between both formulations is apparent where
in particular $-\partial\phi^{(*)}/\partial t\Leftrightarrow\psi_{e/m}$
and therefore scalarization of \eqref{eq:29}\eqref{eq:30} might
also be possible. However whereas for a uniaxial medium the functions
$\phi^{(*)}$ or $\psi_{e/m}$ are sufficient for obtaining scalar
equations, for a gyrotropic medium the additional freedom offered
by the gauge function $G_{c}$ in \eqref{eq:95} (and $L_{c}$) is
needed and these functions do not occur in \eqref{eq:29}\eqref{eq:30}.
With some hindsight we realize that transformations of the {}``electric''
and {}``magnetic'' stream potentials according to\begin{eqnarray}
\epsilon^{-1}\cdot\overline{p}' & = & \epsilon^{-1}\cdot\overline{p}-\overline{\mathcal{E}}\label{eq:105}\\
\epsilon^{-1}\cdot\overline{p}'^{*} & = & \epsilon^{-1}\cdot\overline{p}^{*}+\overline{\mathcal{E}}\label{eq:107-1}\\
\overline{m}' & = & \overline{m}+\overline{\mathcal{H}}\label{eq:106}\\
\overline{m}'^{*} & = & \overline{m}^{*}-\overline{\mathcal{H}}\label{eq:108}\end{eqnarray}

correspond with the following transformations of the current densities

\begin{eqnarray}
\overline{J}' & = & \overline{J}+\nabla\times\overline{\mathcal{H}}-\epsilon\cdot\frac{\partial\overline{\mathcal{E}}}{\partial t}\label{eq:102}\\
\overline{J}'^{*} & = & \overline{J}^{*}-\mu\cdot\frac{\partial\overline{\mathcal{H}}}{\partial t}-\nabla\times\overline{\mathcal{E}}\label{eq:103}\end{eqnarray}

These transformations are thus legitimate provided the fields are
transformed according to\begin{equation}
\overline{E}'=\overline{E}+\overline{\mathcal{E}}\qquad\overline{H}'=\overline{H}+\overline{\mathcal{H}}\label{eq:104}\end{equation}

With the additional transformation \eqref{eq:102}\eqref{eq:103}
of the current densities the formulations based on the standard vector
potentials on one hand and on the Hertz vectors on the other hand
become fully equivalent. The former can thus also be scalarized with
a proper choice of the derivatives of the potentials $\partial\phi^{(*)}/\partial t$
and of the longitudinal fields $\mathcal{E}\overline{c}$ and $\mathcal{H}\overline{c}$.
As a result the scalar equations \eqref{eq:100}\eqref{eq:101} also
hold for the vector potentials with the proper substitutions $A_{c}^{(*)}\Leftrightarrow\Pi_{(e/m)c}$,
$p_{c}\Leftrightarrow J_{c}$ and $\mu_{//}m_{c}^{*}\Leftrightarrow J_{c}^{*}$
(omitting the accents). Also for this transformation $\nabla_{\perp}\times\overline{E}_{\perp}$
and $\nabla_{\perp}\times\overline{H}_{\perp}$ are invariant, explaining
why it is also automatically included using the scalar Hertz potentials.

\section{Conclusions}

We have compared three {}``potential'' methods for solving Maxwell's
equations with arbitrary sources in a lineair stratified gyrotropic
medium where the longitudinal symmetry axis (sometimes referred to
as the distinguished axis) is perpendicular to the strata. In particular
we studied the reduction of Maxwell's equations to two scalar equations
(scalarization). A prerequisite for scalarization to be possible and
which we have accepted without proof, is that the constitutive tensors
should not introduce mixing between longitudinal and transversal field
components. A second condition which was often required in the derivations
is that the transversal constitutive tensors should not depend on
the transverse coordinates. It is perhaps interesting to note that
there are no restrictions on the position dependence of the longitudinal
properties $\epsilon_{//}$ and $\mu_{//}$.

Introducing the conventional \emph{vector/scalar potentials} or the
\emph{Hertz vectors} two uncoupled vector equations are obtained.
We have shown that with the limitations already mentioned these equations
can always be scalarized. First the sources must be scalarized: using
the equivalence between electric and magnetic sources and gauge transformations
for the polarizations (stream potentials) the current densities can
always be replaced by unique current densities along the distinguished
axis. The sources for the Hertz vector equations (the stream potentials)
can be scalarized with the gauge transformations only but it is easier
to use \eqref{eq:55} and the scalarized current densities. At this
stage the problem cannot yet be reduced to two scalar differential
equations because there is still cross-coupling between the longitudinal
and transversal components of the vector potentials or Hertz vectors.
However using gauge transformations (see \eqref{eq:33-1},\eqref{eq:50}
and \eqref{eq:95}) these cross-coupling terms can always be eliminated
and two scalar equations are obtained. When using the vector potentials
then, for a gyrotropic medium, an additional transformation between
electric/magnetic sources must be performed (see \eqref{eq:105}-\eqref{eq:104})
and for such a gyrotropic medium the final scalar equations are also
coupled. The two methods are found to be fully equivalent and the
{}``vector potential'' quantities are merely the time derivatives
of the corresponding {}``Hertz vector'' quantities.

We noticed that the appropriate electric/magnetic source transformations
leave the longitudinal {}``total'' current densities $\nabla_{\perp}\times\overline{E}_{\perp}$
and $\nabla_{\perp}\times\overline{H}_{\perp}$ invariant, and this
obviously holds also for the gauge transformations. In a third method,
4 \emph{scalar Hertz potentials} are defined based on Helmholtz decompositions
\eqref{eq:57}\eqref{eq:58} in the transversal plane and it follows
then immediately that 2 of these scalar Hertz potentials, $\Psi$
and $\Theta$, are also invariant under all transformations needed
to obtain scalarization. These are the natural potentials for obtaining
scalarization and this invariance explains why, when using these scalar
Hertz potentials, the scalarized current densities emerge effortlessly.
These scalar Hertz potentials are also equivalent with the longitudinal
components of the scalarized traditional Hertz vectors used in the
2nd method and eventually we conclude that the three methods are fully
equivalent. It remains to be investigated whether this conclusion
still holds for the more complex media for which scalarization has
been obtained using the scalar Hertz potentials \cite{Weiglhofer:1999p4155}.

\bibliographystyle{unsrt}

\end{document}